\documentclass[fleqn,usenatbib]{mnras}
\usepackage{newtxtext,newtxmath}
\usepackage[T1]{fontenc}
\usepackage[utf8]{inputenc}
\usepackage{natbib}
\usepackage{graphicx}
\usepackage{mathtools}
\usepackage{epsfig}
\usepackage{natbib}
\usepackage{multirow}
\usepackage{rotating}
\usepackage{longtable}
\usepackage{verbatim}
\usepackage{url}
\usepackage{hyperref}
\usepackage{CJK}
\usepackage{caption} 
\usepackage{subcaption} 
\usepackage{booktabs}
\usepackage{array}

\DeclareRobustCommand{\VAN}[3]{#2}
\let\VANthebibliography\thebibliography
\def\thebibliography{\DeclareRobustCommand{\VAN}[3]{##3}\VANthebibliography}

\title[Exoplanet Orbital Distribution II]{Exoplanet Orbital Distribution around FGK Sun-$\odot$-like Host Stars II\\ \large a valley in the orbital semi-major axis distribution of sub-Neptunes}

\author[L. Zeng et al.]{Li Zeng,$^{1}$\thanks{E-mail: li.zeng@geo.uio.no, astrozeng@gmail.com}
Stephanie C. Werner,$^{1}$
Stein B. Jacobsen,$^{2}$
Elena Mamonova,$^{1}$
Reidar G. Trønnes,$^{3,1}$
Ramon Brasser,$^{4,1}$
\\
$^{1}$Centre for Planetary Habitability (PHAB), University of Oslo, 0315 Oslo, Norway\\
$^{2}$Department of Earth and Planetary Sciences, Harvard University, Cambridge, MA 02138\\
$^{3}$Natural History Museum, University of Oslo, Sars gate 1, 0562 Oslo, Norway\\
$^{4}$Konkoly Observatory, HUN-REN CSFK, MTA Centre of Excellence; Konkoly Thege Miklos St. 15-17, H-1121 Budapest, Hungary
}

\date{Accepted XXX. Received YYY; in original form ZZZ}

\pubyear{\the\year{}}

\begin{document}
\label{firstpage}
\pagerange{\pageref{firstpage}--\pageref{lastpage}}
\maketitle

\begin{abstract}
    More than one hundred years ago, physics has been revolutionized when people realized that electronic orbitals, or electro-magnetic interactions in general, are quantized. Now, in this study, we are presenting evidence of quantization of planet orbits around stars. Confining a wave in spatial dimensions "quantizes" its wave number. Therefore, this study points to the evidence of the existence of long-range standing waves in the proto-planetary disks. Such waves, although being on a much larger scale of few tens of AU, have already been found by ALMA observation---so called ring-like structure. Now we see that it may exist within 1 AU, and may exert its effect on the existence and distribution of planets within this distance range to the host star. Careful analysis has been carried out to compare the results of different surveys. 
\end{abstract}

\begin{keywords}
exoplanet -- sub-Neptune -- orbit
\end{keywords}

\section{Introduction}

    The orbital distribution of exoplanets around various types of host stars has been the subject of research investigation in recent years~\citep{Gillis2026, Venturini2024, Luque2022, Dattilo2023, Dattilo2024}. There has been suggestion that the bi-modal radius distribution of small planets around G,K-type host stars shifts weights for smaller host stars which include early M-dwarfs and then mid-to-late M-dwarfs, in the sense that the second peak of sub-Neptunes diminishes relative to the first peak of super-Earths towards smaller host stars. This ultimately ties to the nature and origin of planets which make up the second peak, i.e. sub-Neptunes~\citep{Fulton2017ThePlanets, Fulton2018,  VanEylen2017AnRocky, LPSC2017:Zeng2017PlanetFormation, Zeng2017RNAAS, Zeng2018SurvivalMNRAS, PNAS:Zeng2019, Loyd2020, Luque2022, Owen2018AtmosphericExoplanets, Owen2017ThePlanets, Ginzburg2017Core-poweredExoplanets, Cherubim2023, Werlen2025, Holmberg2024, Burn2024}.

    For this study, we carefully select a group of discovered exoplanets in a well-chosen parameter range of 2.1-3.1 Earth radii (R$_{\oplus}$) for exoplanets and 0.6-1.05 solar masses (M$_{\odot}$) for host stars. This radius range corresponds to the so-called sub-Neptunes, whose population peak around 2.6 Earth radii. There is an ongoing debate about whether these sub-Neptunes are water-worlds with a significant H2O-component or gas-dwarfs with rocky cores surrounded by gas envelopes. Without any solar system analogues, the sub-Neptunes represent an intriguing mystery. 
    
    The chosen 2.1-3.1 Earth radius range is motivated by the relatively high abundance of known exoplanets within this range. This range leverages our empirical understanding of exoplanet mass-radius and radius-period distributions from Kepler/TESS data, targeting the bi-modal peak (1.5–4 R$_{\oplus}$) where super-Earths and sub-Neptunes dominate, avoiding sparse extremes like ultra-short-period or Jovian giants.

    

    The selected host star range of 0.6-1.05 solar masses, corresponds to the K- and G-type stars. The G-type main-sequence stars, including our Sun, are characterized by surface temperatures of 5300-6000 K and a range of 0.8-1.05 solar masses. The smaller K-type main-sequence stars cover surface temperatures of 4000-5300 K and a range of 0.6-0.8 solar masses.

    With our selected population of exoplanets orbiting G- and K-type stars. we perform statistical analysis on the orbital semi-major axes of the sub-Neptunes. Interestingly, there is a valley, and possibly additional peaks and valleys, beyond the primary peak (peak 1, i.e., the innermost peak, which is bounded by the inner edge of the distribution on one side) in the probability density function (pdf) of orbital semi-major axis distribution, which is imprinted upon the overall log-uniform distribution in orbital semi-major axis. 
    
    This valley, and possibly additional valleys and peaks, in the pdf of the orbital semi-major axis is the focus of this study. According to the Kepler survey data, this valley (1) exists approximately around $\sim$0.15-0.20 astronimcal units (AU), with possibly an additional peak (2) at $\sim$0.22-0.24 AU, followed by another valley (2) at $\sim$0.30-0.35 AU, followed by another peak (3) at $\sim$0.37-0.40 AU. There are even hints indicating the existences of possible further peaks located at $\sim$0.6 AU (peak 4), $\sim$0.8 AU (peak 5), and $\sim$1.0 AU (peak 6). Although due to small number statistics, we cannot be sure of these further peaks. Together, this makes up an oscillatory behavior in the orbital semi-major axes of sub-Neptunes. 

    

    The origin of such an oscillatory distribution of sub-Neptinian orbits might be related to the ring-like structures observed by the Atacama Large Millimetre Array (ALMA) in several protoplanetary disks surrounding young stars~\citep{Facchini2024, Liu2019, Liu2017, Vioque2025}. 
    
    Then, the question becomes what physical mechanism could have created those ring-structure in the disk within 1 AU. One possibility of a physical mechanism is that a proto-planetary disk is a continuous medium which can support pressure wave propagating within it, and the reflection of such wave at the inner boundary (inner edge) and the outer boundary can create \emph{standing wave} and \emph{discretize} the wave number. This long-range standing wave, if it ever exists in the disk, would leave its imprints on the orbital distribution of planets formed from it. 

    More than one hundred years ago, physics has been revolutionized when people realized that electronic orbitals, or electro-magnetic interactions in general, are quantized. Now, in this study, we are presenting evidence of quantization of sub-Neptune orbits around G,K-type host stars.


\section{Selection of planets and host stars}

Our previous work (Manuscript I~\citep{Zeng2026a}) has shown that the orbital separation distribution of the small exoplanet population below 4 Earth radii follows approximately a logarithmic-uniform distribution in semi-major axis. However, we also recognize that planets in between the sizes of 1 to 4 Earth radii exhibit a huge diversity in terms of their masses and internal compositions, according to our somewhat limited mass measurements from the synthesis of multiple ground-based radial-velocity measurements of planet masses, and also multiple attempts of spectroscopic characterization of planet atmospheres of such sub-Neptunes from space missions such as JWST. Therefore, this population of 1 to 4 Earth radii, or even a narrower range of 2-4 Earth radii, may still be an intermingled population from various sources and various formation mechanisms. 

\begin{figure*}
\centering
\includegraphics[width=\textwidth]{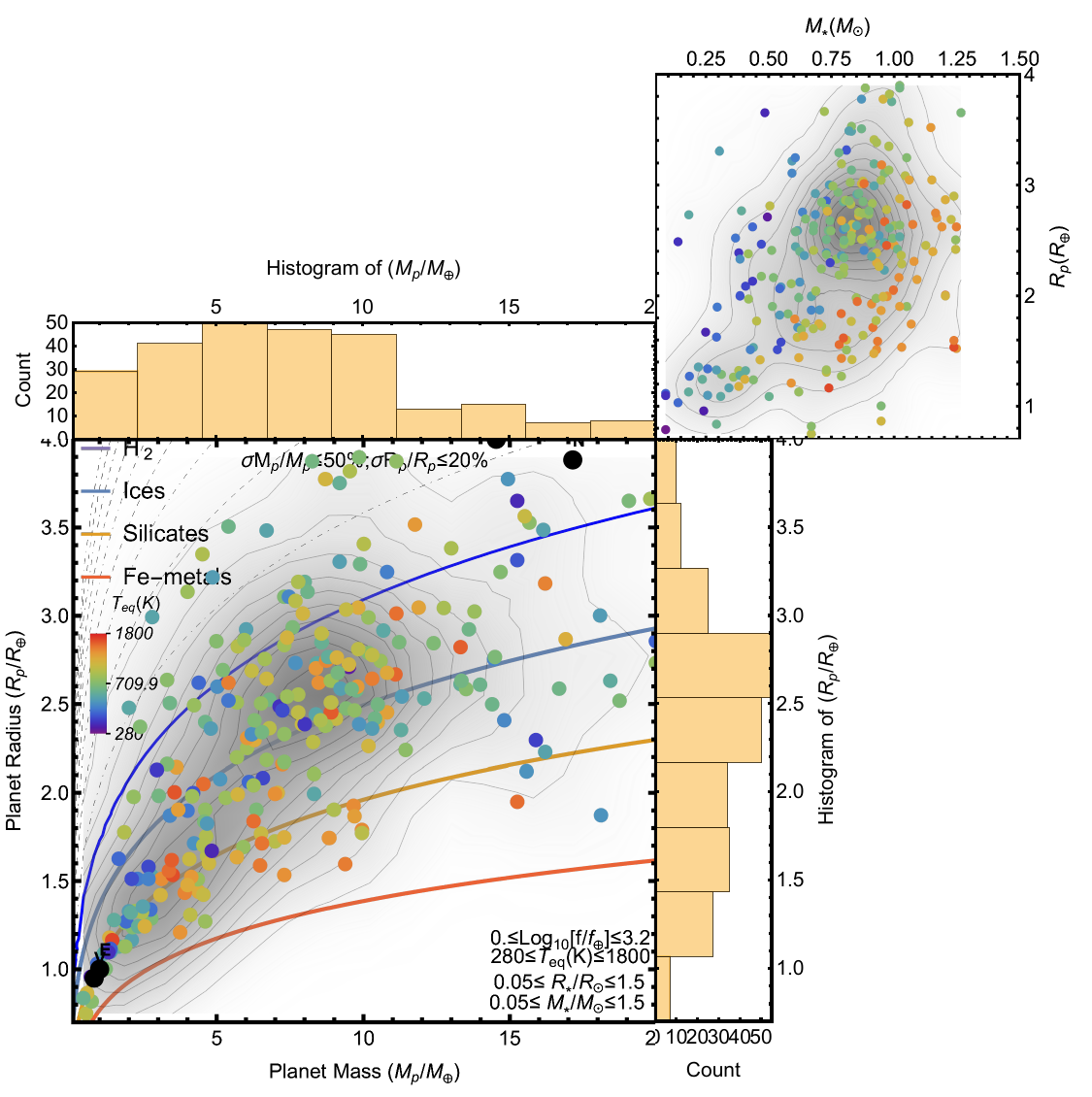}
\caption{Main plot: mass-radius plot of well-studied exoplanets from TepCat in between 1-20 Earth masses and 1-4 Earth radii. Some theoretical mass-radius curves are shown for comparison. Although a subset of planets is below the silicate (i.e. pure rock) curve and can be identified as rocky planets, a larger proportion of the planets is above this curve.  In particular, they exhibit a clustering in the range of 6-10 Earth masses as shown by the background with 2D probability density contours and shading. 
}
\label{fig:mr_plot}
\end{figure*}
\begin{figure}
\centering
\includegraphics[scale=1.0]{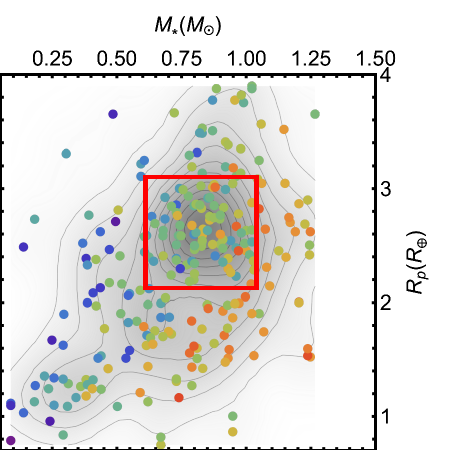}
\caption{Distribution of exoplanet radius (Rp/R-Earth) versus stellar mass (M-star/M-sun), with 2D probability density contours and shading. The planet symbols, colour-coded for equilibrium temperature, are as in Fig.~\ref{fig:mr_plot}. The centre of the sub-Neptunian cluster is at approximately 2.6 times the Earth's radius with a host star mass of about 0.825 times the solar mass. The thin red rectangle outlines a smaller selection of sub-Neptunian planets from the central cluster. This planeary selection will be used in our discussion below. }
\label{fig:mstar_rplanet_plot}
\end{figure}

In Figures~\ref{fig:mr_plot} and ~\ref{fig:mstar_rplanet_plot} we use the mass-radius distribution for exoplanets in the ranges 1-20 Earth masses and 1-4 Earth radii to identify the center of the population in terms of the exoplanet mass versus radius distribution (ratios of M-planet/M-Earth versus R-planet/R-Earth), as well as the the host star mass (M-star/M-Sun) versus exoplanet radius relations. Here we draw planet data from the TepCat~\citep{TEPCat2011JohnSouthworth} database available at \url{https://www.astro.keele.ac.uk/jkt/tepcat/allplanets-ascii.txt}.

In the main plot of Figure~\ref{fig:mr_plot}, we demonstrate the clustering of sub-Neptunes around 7-10 Earth masses and 2.4-2.8 Earth radii, which are delineated by the 2-dimensional (2D) probability density contours and shadings in the background. They also show this range in the (1D) histograms for mass or radius distribution, separately. The formation pathway leading to this clustering and the ubiquity of sub-Neptune planets within the 6-11 Earth mass range, orbiting Sun-like G- and K-type stars, will be discussed in Section~\ref{Discussion}.



\begin{figure}
\centering
\includegraphics[scale=1.0]{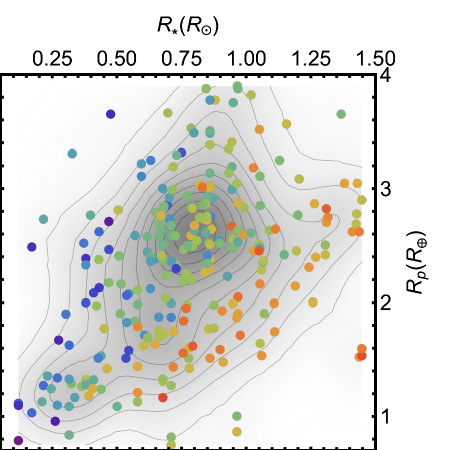}
\caption{Distribution of exoplanet radius (Rp/R-Earth) versus stellar radius (R-star/R-sun), with 2D probability density contours and shading. The planet symbols, colour-coded for equilibrium temperature, are as in Fig.~\ref{fig:mr_plot}. By comparison with the Fig.~\ref{fig:mstar_rplanet_plot}, we decide to adopt stellar mass as a better criterion in selecting their planets rather than stellar radii. }
\label{fig:rstar_rplanet_plot}
\end{figure}

The clustering of sub-Neptunes, also in terms of planetary radius versus host star mass and planetary radius versus host star radius is cleary seen in Figs.~\ref{fig:mstar_rplanet_plot} and~\ref{fig:rstar_rplanet_plot}, respectively. The centre point of these clusters is at a planetary radius of about 2.6 times the Earth's radius and at a host star mass and host star radius of about 0.8 times the solar mass. The clustering in the host star mass or radius versus planetary radius relations in Figs.~\ref{fig:mstar_rplanet_plot} and~\ref{fig:rstar_rplanet_plot} is tighter with a rounder (more equidimensional) distribution in comparison with the more elongated distribution in the planetary radius versus planetary mass diagram of Fig.~\ref{fig:mr_plot}. The selected cluster of host stars and planets within the thin red rectangle in Fig.~\ref{fig:mr_plot} are characterized by a mean star mass of (0.60+1.05)/2 = 0,825 M(solar) and a mean planetary radius of (2.1+3.1)/2 = 2.60 R(Earth).  Below, we will focus on this selected mass-size range.

Stellar mass is the fundamental parameter of a star; virtually all other properties evolve and change over time. This argument should strengthen our choice.

Now, let us turn our attention to the Kepler survey results available at NASA Exoplanet Archive~\citep{Akeson2013TheResearch}~\citep{Christiansen2025}~\citep{Thompson2017Planetary25}~\url{https://exoplanetarchive.ipac.caltech.edu/}. More specifically, we focus on the "Q1-Q17 DR25 Supp Done", which refers to the completion of the supplemental Kepler Object of Interest (KOI) activity table for Data Release 25 (DR25), covering quarters 1 through 17. The reason for this choice of dataset is that most exoplanets do not have mass measurements, and in order to gain a meaningful insight into their orbital separation distribution, we can only rely upon the results of a long-term systematic survey as obtained by the Kepler mission, which were observing the same patch of sky and same stars consistently for three years. We will later on compare this result deduced from \emph{Kepler} to that deduced from \emph{TESS}. 

We also remove the data entry deemed "FALSE POSITIVE" in the sense that we do not keep the data entry denoted "FALSE POSITIVE" by either the Exoplanet Archive Disposition or Disposition using Kepler Data, but the keep the ones marked "CONFIRMED" or "CANDIDATE". Then, we select planets with their host stars in the parameter ranges of 2.1-3.1 Earth radii (R$_{\oplus}$) and 0.6-1.05 solar masses (M$_{\odot}$). We are left with 798 planets in the Kepler's Q1-Q17 sample, after the selection is complete. 

\section{Analysis}

\subsection{Complementary Cumulative Distribution Function}

\begin{figure}
\centering
\includegraphics[width=\columnwidth]{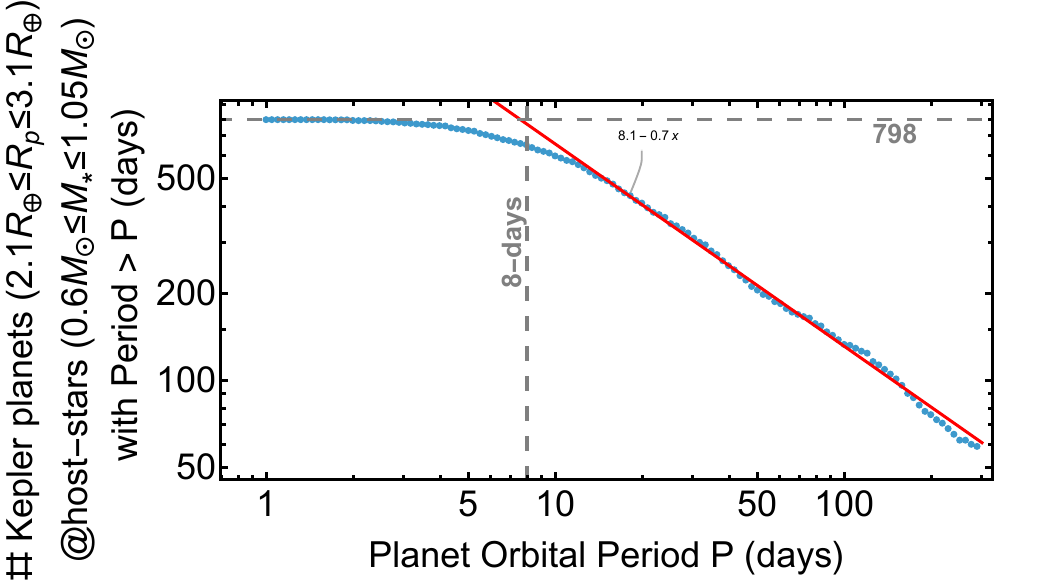}
\caption{Survival Function of Planet Orbital Period Distribution (\emph{P} in days), for selected Kepler planet candidates, in a well-chosen parameter range of 2.1-3.1 R$_{\oplus}$ and 0.6-1.05 M$_{\odot}$, 798 in total, from the Kepler Q1-Q17 DR25 out of the NASA Exoplanet Archive~\citep{Akeson2013TheResearch}~\citep{Christiansen2025}~\citep{Thompson2017Planetary25}. X-axis is the orbital period \emph{P}. Y-axis is the number of \emph{Kepler} planet candidates with orbital period larger than a given \emph{P}. The best fit to the data in the 15-200 day orbital period range is a power-law in logarithmic scale with power index of approximately $-0.7$ shown as the red line. Here, 8-day can be regarded as some sort of inner cut-off in \emph{P} for this group of exoplanets.}
\label{fig:survival_P}
\end{figure}

Fig.~\ref{fig:survival_P} shows the un-corrected \emph{complementary Cumulative Distribution Function} (\textbf{cCDF}), known as the \emph{Survival Function} (\textbf{SF}), of the orbital period \emph{P}, where "un-corrected" stands for "un-corrected of geometric transit probability" as we have discussed in Manuscript I~\citep{Zeng2026a}.

\begin{figure}
\centering
\includegraphics[width=\columnwidth]{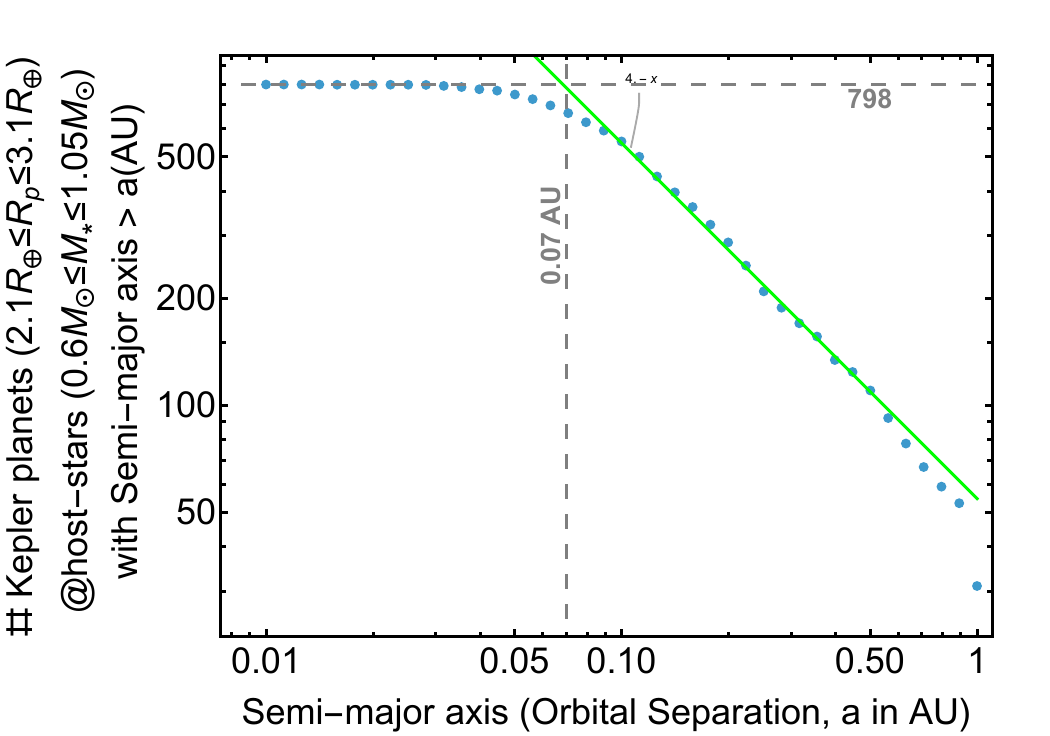}
\caption{Survival Function of Planet Semi-major Axis Distribution (\emph{a} is measured in \emph{astronomical unit} (AU)), for selected Kepler planet candidates, in a well-chosen parameter range of 2.1-3.1 R$_{\oplus}$ and 0.6-1.05 M$_{\odot}$, 798 in total, from the Kepler Q1-Q17 DR25 out of the NASA Exoplanet Archive~\citep{Akeson2013TheResearch}~\citep{Christiansen2025}~\citep{Thompson2017Planetary25}. X-axis is the semi-major axis \emph{a}. Y-axis is the number of \emph{Kepler} planet candidates with semi-major axis larger than a given \emph{a}. The best fit to the data in the 0.1-0.5 AU range is a power-law in logarithmic scale with power index of approximately $-1$ shown as the green line. Here, 0.07 AU can be regarded as some sort of inner cut-off in \emph{a} for this group of exoplanets.}
\label{fig:survival_a}
\end{figure}

Fig.~\ref{fig:survival_a} shows the un-corrected \textbf{SF} of the orbital semi-major axis \textbf{a}, where "un-corrected" stands for "un-corrected of geometric transit probability" as we have discussed in Manuscript I~\citep{Zeng2026a}.

Fig.~\ref{fig:survival_P} and Fig.~\ref{fig:survival_a} already start to show some sort of oscillatory behavior on top of the linear slope fits. This oscillatory behavior is the topic of discussion in the next section. 

Before proceeding to that, we could also use Fig.~\ref{fig:survival_P} and Fig.~\ref{fig:survival_a} to estimate the average (smoothed) occurrence rate of such planets in the radius range of 2.1-3.1 Earth radii, around host stars in the mass range of 0.6-1.05 solar masses. Now, let us look at the Kepler target star sample in Table~\ref{tab:kepler_target_star_table}.

\begin{table}
    \centering
    \caption{\textit{Kepler target stars}~excerpt from~\protect\cite{Batalha2009}. It shows the Kepler target stars divided in various effective temperature bins as well as in visual magnitude bins.}
    \label{tab:kepler_target_star_table}

    \resizebox{\columnwidth}{!}{
    \begin{tabular}{c *{9}{r}}
        \toprule
        Mag/$T_{\mathrm{e}}$(K) & 10500 & 9500 & 8500 & 7500 & 6500 & 5500 & 4500 & 3500 & Total \\
        \midrule
        7.5 & 2 & 8 & 8 & 8 & 8 & 7 & 0 & 0 & 41 \\
        8.5 & 8 & 20 & 26 & 24 & 50 & 16 & 7 & 8 & 159 \\
        9.5 & 9 & 31 & 81 & 65 & 117 & 88 & 11 & 4 & 406 \\
        10.5 & 27 & 37 & 100 & 209 & 405 & 362 & 40 & 9 & 1189 \\
        11.5 & 24 & 58 & 172 & 396 & 1495 & 1356 & 157 & 39 & 3697 \\
        12.5 & 33 & 43 & 230 & 678 & 4148 & 4761 & 625 & 62 & 10580 \\
        13.5 & 34 & 51 & 170 & 737 & 9250 & 15841 & 2218 & 159 & 28460 \\
        14.5 & 3 & 0 & 0 & 0 & 4791 & 29291 & 4401 & 552 & 39038 \\
        15.5 & 7 & 3 & 0 & 0 & 4261 & 43132 & 11188 & 1828 & 60419 \\
        \midrule
        Total & 147 & 251 & 787 & 2117 & 24525 & 94854 & 18647 & 2661 & 143989 \\
        \bottomrule
    \end{tabular}
    }
\end{table}

So, the mass range of 0.6-1.05 solar masses would roughly correspond to the G-type main-sequence stars (5300-6000 K, and 0.8-1.05 R$_{\oplus}$) and the K-type main-sequence stars (4000-5300 K, and 0.6-0.8 R$_{\oplus}$). The total number of such stars in the Kepler sample (Table~\ref{tab:kepler_target_star_table}) is: 

\begin{equation}
    18467+94854=113501
\end{equation}

The mean mass/radius of these host stars (in solar units) that we are interested in is: 

\begin{equation}
    \frac{(0.6+1.05)}{2} = 0.825
\end{equation}

The conversion between one astronomical unit (AU) and one solar radius (R$_{\oplus}$) is as follows.

\begin{equation}
    \frac{1\text{AU}}{1 R_{\odot}} = \frac{149600000 \text{km}} {696000 \text{km}} \approx 215.
\end{equation}

Therefore, the occurrence rate of such planets in between the radius range of 2.1-3.1 R$_{\oplus}$, around host stars in between the mass range of 0.6-1.05 solar masses, per host star on average, per $\ln a$ (per natural logarithmic spacing in orbital semi-major axis), and after correcting for the geometric transit probability, is: 

\begin{equation}
    \frac{dN_{\text{2.1-3.1}R_{\oplus}}}{d\ln a} \approx \frac{798}{(18467+94854)} \cdot \frac{(0.07\pm0.02) \text{AU}}{0.825 R_{\odot}} \approx 0.13 \pm 0.03
\end{equation}

This result is consistent with the result derived in Manuscript I~\citep{Zeng2026a}. This number is slightly smaller than the occurrence rate for the sub-Neptune planets between 2-4 R$_{\oplus}$ derived there, due to a tighter selection range of 2.1-3.1 R$_{\oplus}$. This radius range of 2.1-3.1 R$_{\oplus}$ also corresponds to the peak of this population according to the analysis in this manuscript. The most typical planet in this population would have a radius of: 

\begin{equation}
    \frac{(2.1+3.1)}{2} = 2.6~R_{\oplus}
\end{equation}

This planet radius of 2.6~R$_{\oplus}$ would correspond to a typical planet mass of 8-9 M$_{\oplus}$ according to the currently known mass-radius distribution of exoplanets (see Fig.~\ref{fig:mr_plot}). Such a planet would mostly likely orbit around a host star of 0.825 M$_{\odot}$ or 0.825 R$_{\odot}$, which lies at the borderline of a K-dwarf/G-dwarf main-sequence star, slightly less massive than that of our own Sun. 

Such a planet would most likely consist of a significant amount of H$_2$O, that is, a water world. It would also contain other elements such as Carbon, Nitrogen, Oxygen, and their compounds in its interior and atmosphere. Such a planet may be formed commonly under favorable conditions in the disk. 

\subsection{Oscillatory behavior in Semi-major Axis Distribution?}\label{PDFina}

\begin{figure}
\centering
\includegraphics[scale=1.0]{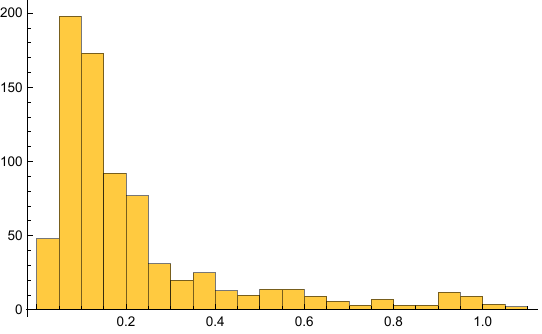}
\caption{Histogram of semi-major axis distribution for 798 planet candidates in our selected parameter range 2.1-3.1 R$_{\oplus}$ and 0.6-1.05 M$_{\odot}$ from the Kepler Q1-Q17 DR25 database~\citep{Akeson2013TheResearch}~\citep{Christiansen2025}~\citep{Thompson2017Planetary25}. X-axis: semi-major axis in AU. Y-axis: number of Kepler planet candidates within each semi-major axis bin, where the bin size is chosen to be 0.050 AU.}
\label{fig:Histogram_a}
\end{figure}

The histogram in Fig.~\ref{fig:Histogram_a} (binned at 0.05 AU) shows the semi-major axes of the 798 selected planets within the Kepler Q1-Q17 DR25 database in the NASA Exoplanet Archive ~\citep{Akeson2013TheResearch}~\citep{Christiansen2025}~\citep{Thompson2017Planetary25}, un-corrected of geometric transit probability as we have discussed in Section~\ref{GeometricTransitProbability}. A first glance shows a primary peak (peak 1, i.e., the innermost peak bounded by the inner edge on one side) located around $\sim$0.1 AU, with the distribution tapering off towards larger orbital separations. Part of the tapering-off is due to the geometric transit probability, which when un-corrected leads to a decrease in the likelihood of seeing planet transit with increasing orbital separations. 

\begin{figure}
\centering
\includegraphics[scale=1.0]{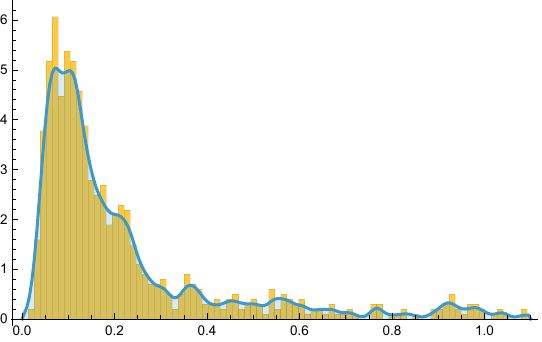}
\caption{Probability Density Function of planet semi-major axis distribution for 798 planet candidates in our selected parameter range 2.1-3.1 R$_{\oplus}$ and 0.6-1.05 M$_{\odot}$ from the Kepler Q1-Q17 DR25 database~\citep{Akeson2013TheResearch}~\citep{Christiansen2025}~\citep{Thompson2017Planetary25}.
X-axis: semi-major axis in AU. Y-axis:  arbitrary unit represents the likelihood of \emph{Kepler} planet candidates within a certain semi-major axis (\emph{a}) range. A Smooth Kernal Distribution fits this histogram with a bandwidth chosen to be 0.013 AU. It indicates the possibility of multiple peaks in the distribution located beyond the primary peak at approximately 0.22, 0.37, 0.57, 0.77, and 0.93 AU. }
\label{fig:SmoothDensityPDF_a}
\end{figure}

\begin{figure}
\centering
\includegraphics[scale=1.0]{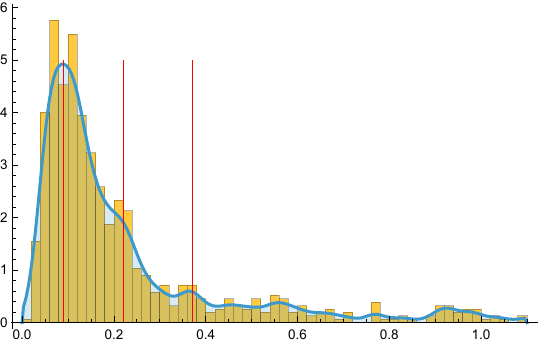}
\caption{Probability Density Function of planet semi-major axis distribution for 798 planet candidates in our selected parameter range 2.1-3.1 R$_{\oplus}$ and 0.6-1.05 M$_{\odot}$ from the Kepler Q1-Q17 DR25 database~\citep{Akeson2013TheResearch}~\citep{Christiansen2025}~\citep{Thompson2017Planetary25}.
X-axis: semi-major axis \emph{a} in AU. Y-axis: arbitrary unit represents the likelihood of \emph{Kepler} planet candidates within a certain semi-major axis (\emph{a}) range. A Smooth Kernal Distribution fits this histogram with a bandwidth chosen to be 0.020 AU. Vertical red lines indicate the location of peak 1,2,3 located at approximately 0.09 AU, 0.22 AU, and 0.37 AU.}
\label{fig:SmoothDensityPDF_a2}
\end{figure}

Based on this histogram (Fig.~\ref{fig:Histogram_a}), we then produce a smooth density histogram (binned at 0.013 AU, Fig.~\ref{fig:SmoothDensityPDF_a}, and binned at 0.020 AU, Fig.~\ref{fig:SmoothDensityPDF_a2}) to show the Probability Density Function (PDF) which approximates this distribution. To our great surprise, this PDF starts to show several peaks and valleys beyond the primary peak, which we call peak 1 located at approximately $\sim$0.1 AU. First of all, there is \emph{hump} (we call it peak 2) located slightly in excess of 0.2 AU at approximately $\sim$0.22 AU. Then, there is another peak (peak 3) located slightly short of 0.4 AU at approximately $\sim$0.37 AU. Then, there are additional peaks located at approximately $\sim$0.57 AU (peak 4), $\sim$0.77 AU (peak 5), and $\sim$0.93 AU (peak 6). Because of small number statistics, we are not so sure about these distant peaks. However, we want to investigate the nature of peak 1 ($\sim$0.09 AU), peak 2 ($\sim0.22$ AU) and peak 3 ($\sim$0.37) and the valleys between them. 

First of all, peak 1 ($\sim$0.09 AU) seems to be a double-spike under finer resolution (See Fig.~\ref{fig:SmoothDensityPDF_a} and Fig.~\ref{fig:SmoothDensityPDF_a2}). That is to say, the primary peak, which is also the innermost peak, could in principle be a composite one consisting of two closely-spaced peaks. 

Secondly, do we see the peak 2 ($\sim0.22$ AU) in other datasets such as TESS (Transiting Exoplanet Survey Satellite) or TepCat? The answer is YES! We will discuss it in the next sub-section. 

\subsection{TESS}

Time we focus on the TESS Project Candidates from the NASA Exoplanet Archive~\citep{Akeson2013TheResearch}~\citep{Christiansen2025}~\citep{Thompson2017Planetary25}~\url{https://exoplanetarchive.ipac.caltech.edu/}. Here we download the TESS Object of Interest (TOI) Table on \date{December 7th, 2025}. After removing all the false positives (FP), it contains 6595 planets. Because the TOI table does not provide the stellar masses but instead only have stellar radii, we are forced to adopt stellar radii as a criterion for further selection in this case for TOI. 

Now, after imposing the criteria of $0.6R_{\odot} \leq R_{\star} \leq 1.05R_{\odot}$ and $2.1R_{\oplus} \leq R_{\text{p}} \leq 3.1R_{\oplus}$, we are left with 425 planets. We implicitly assume that the G-dwarf and K-dwarf main-sequence stars obey an approximately linear mass-radius relation in the sense that the stellar mass is directly proportional to stellar radii. 

\begin{equation}
    M_{\star}/M_{\odot} \approx R_{\star}/R_{\odot}
\end{equation}

Furthermore, because the TOI table itself does not give orbital separation, we have to calculate that from the knowledge of the orbital period ($P$ in days) of planet together with the host stellar mass (here inferred from host stellar radius) by using the Kepler's Third Law. 

\begin{equation}
    a = \left( \frac{M_{\star}}{M_{\odot}} \right) \cdot \left( \frac{P}{365.25} \right)^{2/3} \approx \left( \frac{R_{\star}}{R_{\odot}} \right) \cdot \left( \frac{P}{365.25} \right)^{2/3}
\end{equation}

\begin{figure}
\centering
\includegraphics[scale=1.0]{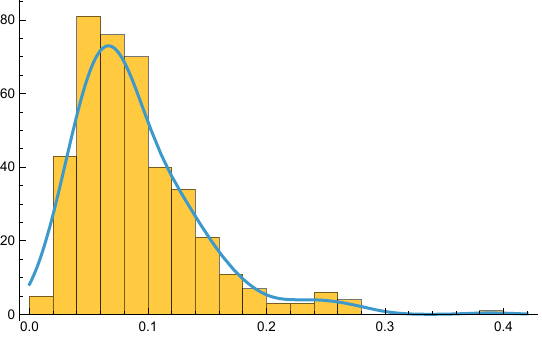}
\caption{Histogram of semi-major axis distribution, for selected TESS planet candidates in a well-chosen parameter range of 2.1-3.1 R$_{\oplus}$ and 0.6-1.05 R$_{\odot}$, 425 in total, from TESS Object of Interest downloaded of the NASA Exoplanet Archive~\citep{Akeson2013TheResearch}~\citep{Christiansen2025}~\citep{Thompson2017Planetary25} as of \date{December 7th, 2025}. X-axis is the orbital separation \emph{a} in AU. Y-axis is the binned number of TESS planet candidates within each orbital separation bin. Here the bin size is chosen to be 0.020 AU. Here, in addition to the primary peak (1) located slightly short of 0.10 AU, we identify a second peak (2) located slightly in excess of 0.20 AU. }
\label{fig:Histogram_a_TESS}
\end{figure}

Please see Fig.~\ref{fig:Histogram_a_TESS}. The primary peak (peak 1) located at approximately 0.1 AU, and the peak 2 located slightly in excess of 0.2 AU, are visible in TESS data. Due to the differences in observational strategies between TESS and Kepler, the Kepler mission was more sensitive to longer orbital-period or larger orbital-separation planets~\cite{Winn2018PlanetSurveys}. Therefore, we do not expect TESS to reproduce the distant peaks beyond $\sim$0.4 AU.

\subsection{TepCat}

\begin{figure*}
\centering
\includegraphics[width=\textwidth]{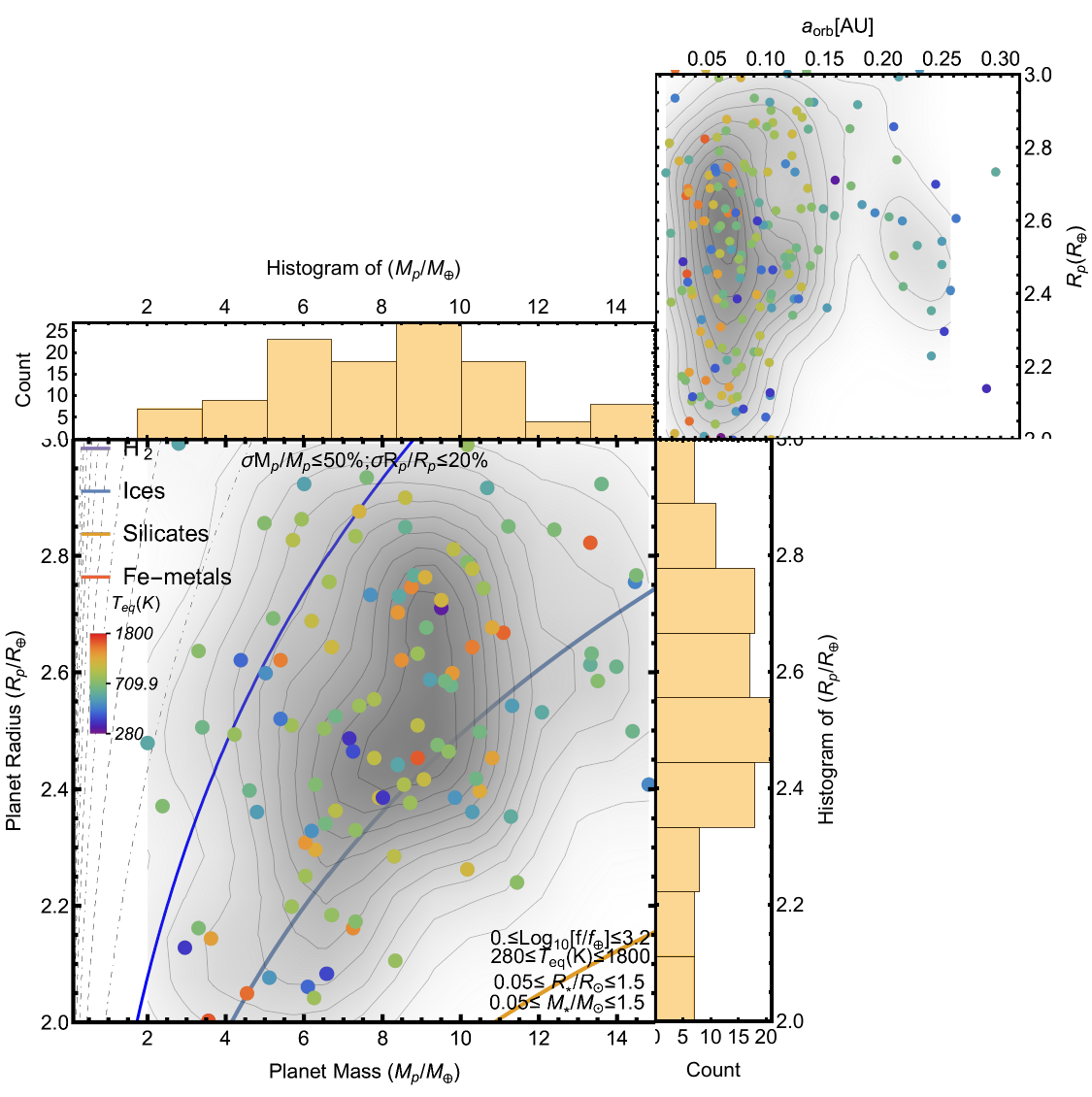}
\caption{Main plot: mass-radius plot of well-studied exoplanets in between 1-15 Earth masses and 2-3 Earth radii. Some theoretical mass-radius curves are shown for comparison. \\
Subplot at the top right corner: Orbital Separation (a)-Planet Radius (R$_{\text{p}}$) plot. The background 2D Histogram with its shadings and contours shows clearly a gap or valley in the planet-star orbital separation in astronomical units (AU). Notice the clustering of planets around 0.05-0.1 AU, which is the primary peak (1), then separated by a valley of low occurrence of planets, until the next peak (2) around 0.2-0.25 AU. }
\label{fig:mr_plot2}
\end{figure*}

The valley or gap in orbital separation (AU) in between the first two peaks (peak 1 and peak 2) in orbital separations could also be seen on the mass-radius plot from the well-studied transiting planets in the TepCat database~\url{https://www.astro.keele.ac.uk/jkt/tepcat/}, which has synthesized all planets with not only radius but also precise mass measurements. Here we zoom in upon the planets in between 2-3 R$_{\oplus}$. Please see Figure~\ref{fig:mr_plot2}, in particular, the subplot at the upper right corner.

\subsection{Further Implications}

\begin{figure}
    \centering
    \begin{subfigure}[b]{0.45\textwidth}
        \centering
        \includegraphics[width=\textwidth]{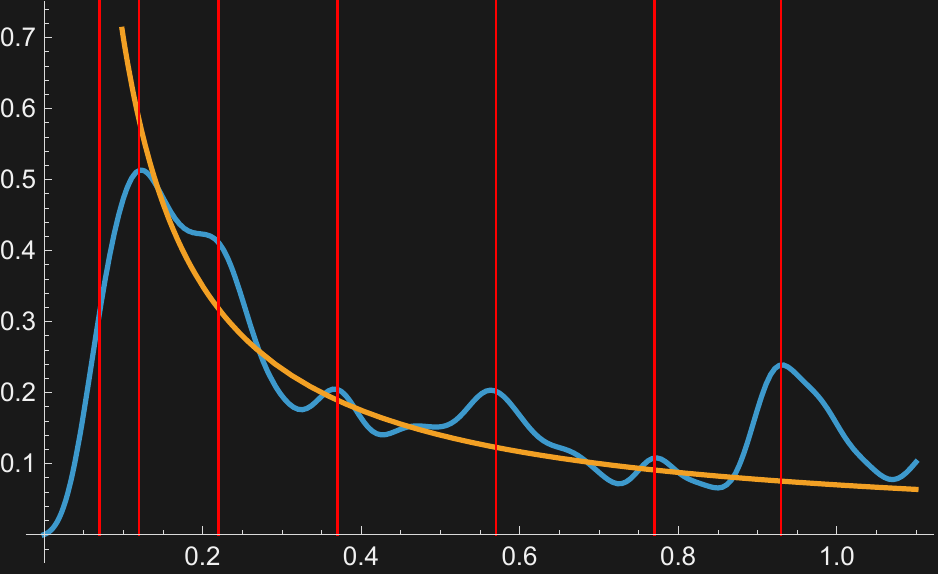} 
        \label{fig:corr1}
    \end{subfigure}
    \hfill 
    \begin{subfigure}[b]{0.45\textwidth}
        \centering
        \includegraphics[width=\textwidth]{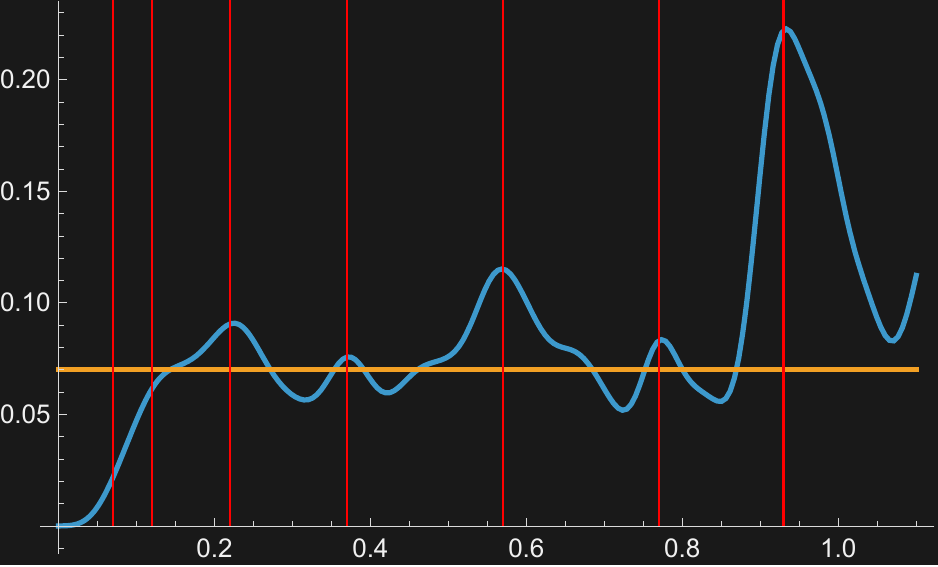} 
        \label{fig:corr2}
    \end{subfigure}
    \caption{Probability density distribution of orbital distances, (a) after correcting for the geometric transit probability, (b) after flattening the background log-uniform distribution (thick orange curve).  The peaks are marked with vertical red lines at 0.07, 0.12, 0.22, 0.37, 0.57, 0.77 and 0.93 AU. The background log-uniform distribution of orbital distances (orange curve) is inversely proportional to the semi-major axis.}
    \label{fig:overall2}
\end{figure}

 After correcting the geometric transit probability, the peaks look like as shown in Fig.~\ref{fig:overall2}. A more crude approximation of this sequence of modes or peaks would be 0.1 AU (likely a double-spike), then followed by approximately 0.2, 0.4, 0.6, 0.8, and 1.0 AU. 

This is a piece of evidence that the ring-like structure observed by ALMA~\citep{Vioque2025, Facchini2024, Liu2019, Liu2017} in proto-planetary disk on the tens of AU-scale \emph{persists} into the innermost region of the disk in sub-AU scale. 


\section{Discussion: The physics behind the peaks and valleys }
\label{Discussion}
What is the physics behind the peaks and valleys in the semi-major-axis distribution of sub-Neptunes? Do planets probe the disk structure? These are the fundamental questions people shall attempt to answer in the future. 

The finding in this manuscript could perhaps be explained in this way. There are long-range standing waves in the inner proto-planetary disk. Such waves have been postulated theoretically for the inner disk within 1 AU~\citep{Chambers2024}. There could also be multiple condensation fronts of water in the proto-planetary disk. 

The disk is a gaseous continuous medium, which has stress/pressure within it, much more like a drumhead, extended across the two spatial dimensions of the disk mid-plane but confined by the host stellar gravity in the vertical direction, and also limited by boundary conditions at the inner edge and perhaps an outer edge as well. 

Plausibly, standing waves or normal modes could be excited on such a drumhead, just as standing waves or normal modes can be excited on a one-dimensional string in tension or on a two-dimensional drumhead in tension.

\begin{figure}
\centering
\includegraphics[width=1.0\columnwidth]{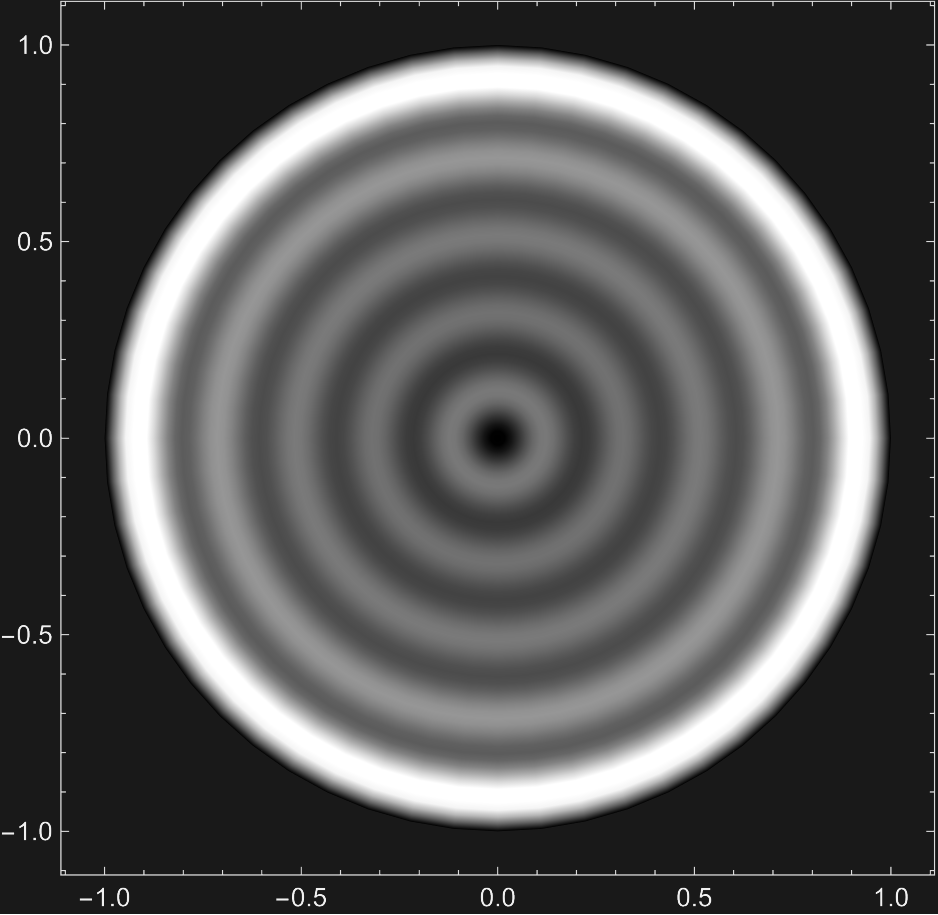}
\caption{Proto-planetary disk Model within 1 AU. }
\label{fig:protoplanetary_disk_1AU}
\end{figure}

These two-dimensional standing waves can be described by cylindrical harmonics or Bessel functions~\citep{Bessel1875} in Mathematical Physics. So we expect in such scenario that there are multiple rings and gaps formed within 1 AU of such a proto-planetary disk. See a schematic drawing in Fig.~\ref{fig:protoplanetary_disk_1AU}. These rings and gaps in turn affect the probability of planets existing at different semi-major axes. 

\section{Conclusion}


We present some evidence of peaks and valleys in the planet semi-major axis distribution of sub-Neptunes around G,K-dwarfs due to unknown physics. This topic shall be explored further and shall have implications for future missions such as PLATO. We seek for collaboration to further explore this matter.

\section*{Acknowledgement}
Li Zeng thanks Trond Reitan on discussion on statistics. This project has received funding from the Research Council of Norway through the Centres of Excellence funding scheme, project number 332523 (PHAB) and project number 360579 (WaterWorlds). Stein B. Jacobsen acknowledges the support from the DOE-NNSA grant DE-NA0004231 to Harvard University: From Z to Planets: Phase V. 

\section*{Data availability}
The data underlying this article are available in \emph{NASA Exoplanet Archive} (~\url{https://exoplanetarchive.ipac.caltech.edu/}). The \emph{Wolfram Mathematica} codes used to process the data are or will become available in Wolfram Community Post, for example, under (~\url{https://community.wolfram.com/groups/-/m/t/3196285}) and (~\url{https://community.wolfram.com/groups/-/m/t/2445247}). The data and codes underlying this article will also be shared on reasonable request to the corresponding author. The authors reserve the rights to modify and improve the codes and data, and to incorporate new data as they become available.

\bibliographystyle{mnras}
\bibliography{mendeley_v7}

\end{document}